\documentclass{article}
\usepackage{spconf,amsmath,graphicx}

\usepackage{enumitem}
\setlist{nosep, leftmargin=14pt}

\usepackage{mwe} %

\title{Transferring Models Trained on Natural Images to 3D MRI via Position Encoded Slice Models}
\name{
\begin{tabular}{c}
  Umang Gupta$^\star$ \qquad
  Tamoghna Chattopadhyay$^\dagger$ \qquad
  Nikhil Dhinagar$^\dagger$ \qquad
  Paul M.\ Thompson$^\dagger$
  \\
  Greg Ver Steeg$^\ddag$\qquad
  The Alzheimer's Disease Neuroimaging Initiative (ADNI)
\end{tabular}
}
\address{
  $^\star$ Information Sciences Institute, University of Southern California \\
  $^\dagger$ Imaging Genetics Center, Mark and Mary Stevens Institute for Neuroimaging and Informatics, \\Keck School of Medicine, University of Southern California\\
  $\ddag$ University of California Riverside
}

\usepackage[utf8]{inputenc} %
\usepackage[T1]{fontenc}    %
\usepackage{hyperref}       %
\usepackage{url}            %
\usepackage{booktabs}       %
\usepackage{amsfonts}       %
\usepackage{nicefrac}       %
\usepackage{microtype}      %
\usepackage{xcolor}         %
\usepackage[]{multirow}
\usepackage[]{amssymb}
\usepackage{mwe}
\usepackage{makecell}
\usepackage{todonotes}
\usepackage{pifont}         %
\usepackage{stfloats}%
\usepackage{amsmath}
\usepackage{arydshln}

\newcommand{\cmark}{\ding{51}}%
\newcommand{\xmark}{\ding{55}}%
\usepackage{lineno}

\newcommand{\four}{10\textsuperscript{-4}}
\newcommand{\three}{10\textsuperscript{-3}}
\newcommand{\five}{10\textsuperscript{-5}}

\newcommand{\adam}{\texttt{ADAM}}
\newcommand{\sgd}{\texttt{SGD}}
\newcommand{\no}{\xmark}
\newcommand{\yes}{\cmark}

\newcommand{\axisone}{sagittal}
\newcommand{\axistwo}{coronal}
\newcommand{\axisthree}{axial}
\newcommand{\result}[2]{#1{ $\pm$ #2}}
\newcommand{\codeurl}{\url{https://github.com/umgupta/2d-slice-set-networks}}

\begin{document}

\maketitle

\begin{abstract}
Transfer learning has remarkably improved computer vision. These advances also promise improvements in neuroimaging, where training set sizes are often small. However, various difficulties arise in directly applying models pretrained on natural images to radiologic images, such as MRIs. In particular, a mismatch in the input space (2D images vs.\ 3D MRIs) restricts the direct transfer of models, often forcing us to consider only a few MRI slices as input. To this end, we leverage the 2D-Slice-CNN architecture of Gupta et al.\ (2021), which embeds all the MRI slices with 2D encoders (neural networks that take 2D image input) and combines them via permutation-invariant layers. With the insight that the pretrained model can serve as the 2D encoder, we initialize the 2D encoder with ImageNet pretrained weights that outperform those initialized and trained from scratch on two neuroimaging tasks --- brain age prediction on the UK Biobank dataset and Alzheimer’s disease detection on the ADNI dataset. Further, we improve the modeling capabilities of 2D-Slice models by incorporating spatial information  through position embeddings, which can improve the performance in some cases.

\end{abstract}


\section{Introduction}\label{sec:intro}

Even though tremendous advances have been made in pretraining computer vision models that work well for natural images such as photographs, it is unclear how easily these pretrained models can be adapted to perform tasks on radiologic images such as MRIs due to domain differences. Moreover, natural images are 2-dimensional (2D), whereas brain MRIs are typically 3-dimensional (3D), making the direct finetuning of the model pretrained on 2D images challenging.
This has led to an active debate in the radiology field on how to pretrain deep learning methods for MRI-based tasks, such as disease classification and staging, identifying pathology, and anatomical  segmentation (e.g., \cite{alzubaidi2021deepening,ke2021chextransfer,valverde2021transfer}).
Many workarounds have been proposed, such as  deriving specialized pretraining datasets such as  YouTube videos~\cite{MALIK2022325} or making predictions from only a few MRI slices~\cite{hon2017towards,valliani2017deep,islam2018brain,MRISignBrainAge}. Using fewer slices severely limits the information available for a machine learning model to make predictions, leading to suboptimal performance.

To this end, we consider recently proposed 2D-Slice-CNN models that can consider full 3D MRI scans as the input. These models process each 2D slice via a slice encoder (usually a 2D CNN) and aggregate the resulting slice embeddings via permutation invariant operations, such as by computing the mean of the embeddings or using self-attention over the embeddings~\cite{gupta2021improved},  max-pooling~\cite{dhinagar2022evaluation}, or RNNs~\cite{lam2020accurate}.
These models can access information from all the slices during training, thus exploiting a richer feature set than models that work with a single slice, and the neural networks pretrained on 2D natural images are excellent candidates to use as the slice encoder. Our first contribution is to study the effect of replacing the slice encoder with a model pretrained on ImageNet.

Our second contribution is incorporating positional encoding in the 2D-Slice model before slice embedding aggregation to preserve spatial information.
The permutation-invariant operations can remove information about the ordering of the slices, thus limiting learning capabilities. Adding positional encoding, i.e., a unique vector corresponding to each position, can help, especially if predictive features are reliably located in specific parts of the images. Positional encoding allows the model to learn spatial information if needed (e.g.,~\cite{SCHLEMPER2019197}).

We extensively evaluate the above-discussed models and a 3D-CNN with comparable architecture for brain age prediction (the common benchmarking task of predicting a person's age from their MRI scan) and an Alzheimer's disease diagnosis (binary classification) task.
Incorporating positional encodings improved test performance in some cases. Contrary to~\cite{gupta2021improved}, we find that 2D-Slice-CNNs perform comparably to 3D-CNN for brain age prediction\footnote{3D-CNN results improved due to expanded hyperparameter search.}. Further, 2D-Slice-CNN models initialized randomly did not perform at par with 3D-CNNs for AD detection. However, the  2D-Slice-CNN models with ImageNet pretrained ResNet-18 encoder outperformed all the models for both tasks, showing that it is possible to transfer inductive biases from natural images to 3D neuroimaging tasks.

\section[2D-slice CNN with Positional Encodings]{2D-slice CNN with Positional Encodings}
\begin{figure*}
    \centering
    {\includegraphics[width=0.85\textwidth, trim={6cm 2.2cm 3cm 8.9cm},clip]{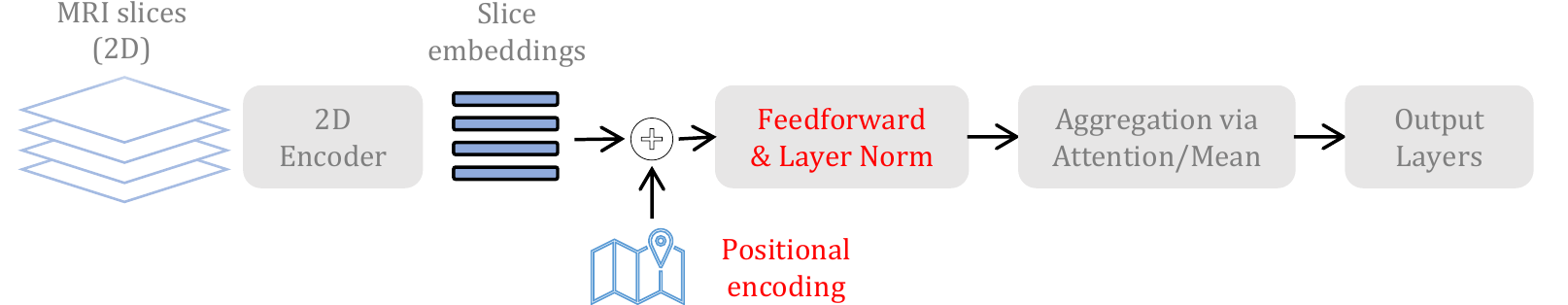}}
    \caption{Positional encoding in the 2D-Slice-CNN architecture. Newer  components (compared to~\cite{gupta2021improved}) are shown in red font.}
    \label{fig:approach}
\end{figure*}

The go-to approach to train models that take raw 3D MRIs as input is to use 3D convolutional layers
(e.g.,~\cite{peng2019accurate, hu2020brain, guanLearning2021}).
Instead,~\cite{gupta2021improved} used 2D-CNNs to encode individual slices and combined the slice representations with permutation-invariant operations such as mean or attention. Compared to 3D-CNNs, these 2D-Slice-based models tend to be less accurate as they may lose spatial information due to permutation invariance. To this end, we encode positional information in the model by introducing positional encodings similar to transformers~\cite{vaswani2017attention,dosovitskiy2021image}. In particular, these are added to the slices' representations and trained end-to-end with other model parameters (see Fig.~\ref{fig:approach}). Suppose the MRI consists of $K$ 2D slices ($x_k, k \in \{1\ldots K\}$) that are each embedded to a d-dimensional representation, $r(x_k)$ with the 2D slice encoder. The model will compute $r(x_k)+p_k$, where $p_k \in \mathbb R^d$ is a trainable vector and depends only on $k$. Fixed positional encodings can also be used. However, this work considers positional encodings as trainable.

Further, motivated by the transformer architecture, we also added a feed-forward layer and layer norm after the attention (i.e., for attention-based aggregation) as that improved the performance on brain age prediction slightly (2.86 reported in~\cite{gupta2021improved} vs.\ 2.83 MAE in our case).

\section{Experiment Setup}

\subsection{Datasets}

We consider two tasks --- Brain age prediction and Alzheimer's disease (AD) detection from brain MRIs.
Our setup for brain age prediction is the same as~\cite{gupta2021improved}.
We used MRI scans from a subset of healthy subjects from the UK Biobank dataset~\cite{ukbb} with no psychiatric diagnosis.  The training, test, and validation set sizes were 7,312; 940; and 2,194; with a mean chronological age and standard deviation of 62.6 and 7.4 years.

We used the same dataset as~\cite{lam20203d} for the AD task. In particular, we use MRI scans from three phases of the  Alzheimer’s Disease Neuroimaging Initiative (ADNI), known as ADNI1,
ADNI2/GO, and ADNI3. These datasets have repeated scans of the same subject. However, the subjects in train/test/validation sets do not overlap. We used
4,561 scans (2,372/873/916 in train/validation/test).  The test set contains 244 scans with AD and 672 with a healthy diagnosis.

All the scans were reoriented to a standard brain MRI template, and the final dimensions of each scan in both datasets were $91\times109\times91$.

\subsection{Model \& Training}
We evaluate the 2D-Slice-CNNs under different settings --- with and without positional encodings and with and without pretrained encoders. We consider two encoder architectures.  We use a 5-layer 2D-CNN  encoder (similar to~\cite{gupta2021improved, lam2020accurate}) adapted from the 3D-CNN baseline of~\cite{peng2019accurate} to benchmark the effect of incorporating positional encodings.

We evaluate the effect of pretraining on natural images (2D) with pretrained ResNets~\cite{he2016deep}. We use ResNets pretrained on ImageNet-1K~\cite{deng2009imagenet} from the PyTorch model hub%
\footnote{\url{https://pytorch.org/hub/}}
as the 2D slice encoder. To get the slice embeddings from ResNets, we remove the final feed-forward layers and consider the output from the last layer (i.e., the average pooling layer). Therefore, the embedding sizes are 512 and 2048 for ResNet-18 and ResNet-50 encoders. For the 5-layer 2D-CNN encoder, the  embedding size is 32. When using pretrained weights, we finetune the entire model, including the encoder. Our implementation is available at \codeurl.

For 2D-Slice  models, we find that slicing MRI along the sagittal axis works best for brain age prediction, as also observed by~\cite{gupta2021improved}. However, this was not the case for the AD task. Therefore we report the results of slicing along all three axes for AD.  We use the same hyperparameters as~\cite{gupta2021improved}, except for the optimizer and learning rate.
We search in \{(\adam, \four), (\sgd, \three / \four / \five)\} for the brain age prediction task.
\adam\ optimizer with a learning rate of \four\ works best for all the models except for the 3D-CNN. We report results for all models on the AD task with a learning rate of \four\ and \adam\ optimizer\footnote{We evaluated 3D-CNN with \sgd\ on the AD task, but \adam\ worked best.}. All the models are trained for 100 epochs, and the best model was chosen based on performance on the validation set at the end of every epoch.

\begin{table*}[t]
    \centering
    \fontsize{10}{11}\selectfont
    \begin{tabular}[]{l cccc ccc}
        \toprule
        \multicolumn{4}{c}{Method}
        &\multirow{2}{*}{\makecell[tc]{Bal.\\ Accuracy}}
        &\multirow{2}{*}{\makecell[tc]{F1\\Score}}
        &\multirow{2}{*}{\makecell[tc]{Avg.\\ Precision}}
        \\
        \cmidrule(l){1-4}
        Encoder &
        Pos. Enc.  &
        Pretrained &
        Axis &
        \\
        \cmidrule(l){1-1}
        \cmidrule(lr){2-2}
        \cmidrule(lr){3-3}
        \cmidrule(lr){4-4}
        \cmidrule(lr){5-5}
        \cmidrule(lr){6-6}
        \cmidrule(l){7-7}

        3D-CNN                & -    & \no  & -          & \result{88.40}{0.703} & \result{84.44}{1.028} & \result{92.33}{0.492} \\
        \cmidrule(l){1-7}

        2D-CNN (Mean)         & \no  & \no  & \axisone   & \result{87.53}{1.311} & \result{84.40}{1.707} & \textbf{\result{94.86}{0.946}} \\ %
        2D-CNN (Mean)         & \no  & \no  & \axistwo   & \result{85.78}{2.811} & \result{81.38}{3.610} & \result{90.57}{2.976} \\ %
        2D-CNN (Mean)         & \no  & \no  & \axisthree & \result{86.34}{1.445} & \result{82.21}{1.573} & \result{90.81}{0.677} \\ %
        \noalign{\vskip 0.25ex}\cdashline{2-7}\noalign{\vskip 0.5ex}
        2D-CNN (Mean)         & \yes & \no  & \axisone   & \result{87.05}{1.115} & \result{83.40}{0.873} & \result{93.40}{0.956} \\ %
        2D-CNN (Mean)         & \yes & \no  & \axistwo   & \result{85.41}{2.521} & \result{80.62}{2.521} & \result{89.86}{0.771} \\ %
        2D-CNN (Mean)         & \yes & \no  & \axisthree & \result{86.61}{0.942} & \result{81.20}{1.971} & \result{91.44}{2.123} \\ %
        \cmidrule(l){1-7}
        2D-ResNet-18 (Mean)   & \no  & \no  & \axisone   & \result{85.71}{4.834} & \result{80.89}{6.111} & \result{91.12}{2.880} \\ %
        2D-ResNet-18 (Mean)   & \no  & \no  & \axistwo   & \result{85.73}{1.433} & \result{81.30}{1.802} & \result{90.60}{2.544} \\ %
        2D-ResNet-18 (Mean)   & \no  & \no  & \axisthree & \result{84.61}{2.099} & \result{79.21}{2.906} & \result{88.95}{1.741}   \\ %
        \noalign{\vskip 0.25ex}\cdashline{2-7}\noalign{\vskip 0.5ex}
        2D-ResNet-18 (Mean)   & \yes & \no  & \axisone   & \result{86.57}{2.871} & \result{81.69}{3.972} & \result{91.80}{2.150} \\ %
        2D-ResNet-18 (Mean)   & \yes & \no  & \axistwo   & \result{84.34}{1.706} & \result{79.75}{2.258} & \result{90.29}{0.645} \\ %
        2D-ResNet-18 (Mean)   & \yes & \no  & \axisthree & \result{86.55}{1.059} & \result{81.39}{1.752} & \result{90.82}{0.701} \\ %
        \cmidrule(l){1-7}
        2D-ResNet-18 (Mean)   & \no  & \yes & \axisone   & \result{87.09}{1.192} & \result{82.07}{1.167} & \result{90.80}{1.601} \\ %
        2D-ResNet-18 (Mean)   & \no  & \yes & \axistwo   & \result{87.31}{0.971} & \result{83.19}{1.518} & \result{91.80}{1.626} \\ %
        2D-ResNet-18 (Mean)   & \no  & \yes & \axisthree & \textbf{\result{88.59}{1.187}} & \textbf{\result{85.10}{1.265}} & \result{93.11}{0.482} \\ %
        \noalign{\vskip 0.25ex}\cdashline{2-7}\noalign{\vskip 0.5ex}
        2D-ResNet-18 (Mean)   & \yes & \yes & \axisone   & \result{87.19}{2.037} & \result{83.16}{2.158} & \result{91.75}{1.291} \\ %
        2D-ResNet-18 (Mean)   & \yes & \yes & \axistwo   & \result{86.95}{2.750} & \result{82.33}{2.811} & \result{90.96}{1.541} \\ %
        2D-ResNet-18 (Mean)   & \yes & \yes & \axisthree & \textbf{\result{88.60}{2.058}} & \result{84.52}{2.389} & \result{92.56}{0.563} \\ %
        \bottomrule
    \end{tabular}
    \caption{Test set results for AD detection (binary labels) on the ADNI dataset. Higher is better for all metrics. Model selection was performed based on balanced accuracy. Results (mean and standard deviations) are reported over 5 runs with different seeds. The first column describes the neural network architecture or the encoder in the case of 2D-Slice models.}
    \label{tab:adni_results}
\end{table*}

\begin{table*}[t]
    \centering
    \fontsize{10}{11}\selectfont
    \begin{tabular}{l cc cc r}
        \toprule
        \multicolumn{3}{c}{Method}
        & \multirowcell{2}{Mean Absolute Err.\\ (MAE)}%
        & \multirowcell{2}{Root Mean Square Err.\\ (RMSE)}%
        & \multirow{2}{*}{\# Params}
        \\
        \cmidrule(l){1-3}
        Encoder &
        Pos. Enc. & Pretrained

        &
        &
        \\
        \cmidrule(l){1-1}
        \cmidrule(lr){2-2}
        \cmidrule(lr){3-3}
        \cmidrule(lr){4-4}
        \cmidrule(lr){5-5}
        \cmidrule(lr){6-6}

        3D-CNN                      & -    & \no  & \result{2.792}{0.032} & \result{3.521}{0.023} &  $\sim$ 3M  \\
        \cmidrule(lr){1-6}
        2D-CNN (Mean)               & \no  & \no  & \result{2.826}{0.021} & \result{3.582}{0.027} &  $\sim$ 1M  \\
        2D-CNN (Mean)               & \yes & \no  & \result{2.847}{0.051} & \result{3.617}{0.067} &  $\sim$ 1M  \\
        2D-CNN (Attention)          & \no  & \no  & \result{2.839}{0.009} & \result{3.590}{0.014} &  $\sim$ 1M  \\
        2D-CNN (Attention)          & \yes & \no  & \result{2.888}{0.067} & \result{3.655}{0.075} &  $\sim$ 1M  \\
        \cmidrule(lr){1-6}
        2D-ResNet-18 (Mean)         & \no  & \no  & \result{2.911}{0.039} & \result{3.684}{0.043} &  $\sim$ 12M \\
        2D-ResNet-18 (Mean)         & \no  & \yes & \textbf{\result{2.715}{0.029}} & \textbf{\result{3.426}{0.044}} &  $\sim$ 12M \\
        2D-ResNet-18 (Mean)         & \yes & \yes & \textbf{\result{2.721}{0.045}} & \textbf{\result{3.439}{0.053}} &  $\sim$ 12M \\
        \cmidrule(lr){1-6}
        2D-ResNet-50 (Mean)         & \no & \yes & \result{2.743}{0.018} & \result{3.468}{0.029} &  $\sim$ 26M \\

        \bottomrule
    \end{tabular}
    \caption{Test set results for the brain age prediction task. Lower MAE and RMSE are better. Model selection was performed based on MAE metric. The first column describes the neural network architecture or the encoder in the case of the 2D-Slice models. All experiments used \axisone\ slices.
    Results (mean and standard deviations) are reported over 5 runs with different seeds.}
    \label{tab:brain_age_results}
\end{table*}

\section{Results}
Tables~\ref{tab:adni_results} and~\ref{tab:brain_age_results}  summarize the performance of different models for  AD and brain age  prediction, respectively. We observed that the 2D-Slice-CNN was better at predicting brain age when using the same hyperparameters as~\cite{gupta2021improved}. However, our 3D-CNN results are better than 2D-Slice-CNN (Table~\ref{tab:brain_age_results}), and the reported numbers in~\cite{gupta2021improved} (3.02 vs.\ 2.792 MAE) due to switching from PyTorch's default initialization to explicitly using He initialization~\cite{he2015delving} and using \sgd\ optimizer. Similarly, 3D-CNN outperforms the 2D-Slice-CNN for AD prediction (88.40 vs.\ 87.53 accuracy). We hypothesized that this might be due to 2D-Slice-CNN combining slice embeddings with permutation invariant operation and consequently losing information about the position of the slice in MRI. We, thus, incorporate positional encodings in the 2D-Slice model.

\subsection{Effect of Position Encodings on the 2D-Slice-CNN.}
Introducing position encodings improved 2D-Slice models for AD prediction in some cases. When using 2D-ResNet-18 encoder (not pretrained), we see improvements in AD prediction when slices along \axisone\ and \axisthree\ directions are used. However, in other cases, the performance on the AD prediction task did not improve when incorporating position encodings.  For brain age prediction, the model performance stayed identical (ResNet Encoders) or deteriorated (5-layer CNN) when using position encodings.
Brain MRIs are usually aligned to a standard template, so the position of specific patterns can provide extra information. However, it may also make the model more sensitive to minor changes in the alignment template and cause overfitting. This could be a potential reason why the performance did not improve on incorporating position encoding.   Here,  we incorporate position encodings in the higher layers. Thus the model may not be able to exploit the spatial information best. Placement of position encodings is crucial~\cite{chen2021demystifying} and left for future work. Overall, position encodings helped in limited cases, and the gains may be task and model dependent.

\subsection{Can Encoders Pretrained on ImageNet improve 3D Deep Neuroimaging?}
We answer this in the affirmative. We finetune 2D-Slice models with ResNet encoders initialized with ImageNet-1K pretrained weights. Initializing with pretrained weights consistently outperforms random initialization (i.e., training from scratch) in all cases (See Tables \ref{tab:adni_results} and~\ref{tab:brain_age_results}). These results validate our hypothesis that a) models trained on natural images (2D) can be helpful for neuroimaging tasks and b) 2D-Slice-CNNs can be used to transfer 2D models to 3D data directly.

Our main goal with these experiments is to demonstrate improvement in performance due to natural image pretraining. Nevertheless, another interesting outcome is that 2D-Slice models with pretrained encoders are the best for both tasks, outperforming the 3D-CNN as well. For the AD  task, we observe that 3D-CNN has a balanced accuracy of 88.40. In contrast, models with pretrained ResNet-18 encoder model have a balanced accuracy of 88.6 when using slices along the axial direction (both with and without position encodings). Similarly, the best MAE with the pretrained model for brain age prediction is 2.715 compared to 2.792 MAE with 3D-CNN. Finally, We also evaluated if increasing the size of the pretrained encoder may lead to more gain by employing ResNet-50 as the encoder for brain age prediction.
We did not see significant improvements over ResNet-18, and we leave further exploration for future work.

\section{Discussion}\label{sec:discussion}

In this paper, we extended the 2D-Slice-based architecture of~\cite{gupta2021improved} by incorporating positional embeddings.we demonstrated improved brain age prediction and AD detection performance by employing slice encoders pretrained on ImageNet-1K, a large dataset of natural 2D images.

Our work contributes to the growing literature on using pretraining for machine learning on radiologic images, for which training datasets are often small. Since most large image datasets are 2D natural images, it is natural to pretrain on natural images.  However, there are two main challenges with this --- a) Domain mismatch, i.e., radiologic images vs.\ natural images; b) Input dimension mismatch, i.e., 2D vs.\ 3D images. Most prior works using off-the-shelf vision models (i.e., pretrained with natural image datasets such as ImageNet) consider a single or a few 2D slices of the MRI scan as the input to avoid the problem of input mismatch~\cite{hon2017towards,valliani2017deep,islam2018brain,MRISignBrainAge}. These may then aggregate the results from different slices during inference only. Such approaches are limiting and lead to suboptimal performance.

In contrast, our approach only substitutes encoders with pretrained counterparts. The model is trained end-to-end and considers the whole MRI as input. It addresses the input dimension mismatch problem without limiting or compromising the information available to the model to make predictions. Despite domain mismatch, pretraining outperforms the models trained from scratch.

Very few datasets exist for pretraining with raw 3D MRI images directly~\cite{chen2019med3d}. Only recently very large radiologic 2D image datasets become have publicly available for pretraining models~\cite{doi:10.1148/ryai.210315,alzubaidi2021mednet}. It would be interesting to pretrain 2D encoders with such datasets in the future to alleviate the domain mismatch problem. In this work, we used models pretrained on the supervised classification task. In future, it would be interesting to evaluate self-supervised pretraining with in- and out-domain images (e.g., \cite{dhinagar2022evaluation}).

\section{Compliance with Ethical Standards}
This is a study of previously collected, anonymized and de-identified data available in a public repository.
\section{Acknowledgements}
We thank the ADNI investigators and their public and private funders for creating and publicly disseminating the ADNI dataset.
This work is supported by NIH (U01AG068057, P01AG055367).

\bibliographystyle{IEEEbib}
{\small\bibliography{refs}}
\end{document}